\documentstyle[12pt]{article}
\begin{document}
\hfill

\centerline{\LARGE  Combinatorial \vspace{5mm}
formulation }
\centerline{\LARGE  of Ising \vspace{10mm} model revisited.}
\centerline{\large G.A.T.F.da Costa \footnote{e-mail: gatcosta@mtm.ufsc.br}
and A. L. Maciel \footnote{Suported by a PIBIC/CNPQ - BIP/UFSC fellowship}  }
\centerline{\large Departamento de Matem\'{a}tica}
\centerline{\large Universidade Federal de Santa Catarina}
\centerline{\large 88040-900- Florian\'{o}polis-SC-\vspace{5mm}Brasil}

\begin{abstract}

In 1952,  Kac and Ward
developed a combinatorial formulation for the two dimensional
Ising model
which is another method of obtaining Onsager's famous formula
for the free energy per site in the termodynamic limit of the model. 
Feynman gave an important contribution to this formulation
 conjecturing a crucial 
mathematical relation 
which completed Kac and Ward ideas.
In this paper, the method of Kac, Ward and Feynman
for the free field Ising model  in two dimensions is reviewed
in a selfcontained way
 and 
Onsager's formula  computed.
 \\
\\
Em 1952, Kac e Ward desenvolveram uma formula\c{c}\~{a}o 
combinatorial do modelo de Ising em duas dimens\~{o}es
que \'{e} um outro m\'{e}todo para se
obter a famosa f\'{o}rmula de Onsager para a energia livre por s\'{\i}tio 
no limite
termodin\^{a}mico do modelo. Feynman fez importante contribui\c{c}\~{a}o a esta
formula\c{c}\~{a}o conjecturando uma rela\c{c}\~{a}o matem\'{a}tica 
crucial que completou
as id\'{e}ias de Kac e Ward. Neste trabalho, o m\'{e}todo de 
Kac, Ward e Feynman para o modelo de Ising em duas dimens\~{o}es sem campo
\'{e} revisada e a f\'{o}rmula de Onsager \'{e} calculada.

\vspace{3cm}
\end{abstract}


\newpage
\section{Introduction.}

The aim of statistical physics is to understand
the macroscopic behaviour of a system formed by a very large number
of particles from information about how they interact with each other.
One way in which one  can gain insight into this problem and thus
about complex systems is by constructing idealized models which hopefully
will exhibit some of the interesting features of real systems like
phase transitions. Perhaps the most studied of these
idealized models is the Ising model so called in honor to
his first investigator, Ernst Ising (1900-1998).

The model was originally proposed as a simple model of ferromagnetism.
In ref. [1] 
Ising investigated the model in one dimension
and computed exactly its partition function.
In 1944, Onsager [2] considered
the free field  model in two dimensions and succeded to compute
the partition function exactly. His method became known as the 
algebraic formulation of the model. In 1952,  Kac and Ward
[3] developed a quite different method of obtaining Onsager results
known as the combinatorial formulation of the Ising model. Feynman
developed the method farther and conjectured an identity relating
functions defined on graphs and functions defined on paths
on a square lattice [4, 7]. This identity is a crucial element in
the combinatorial formulation of  Kac, Ward and Feynman  of the Ising model. 
The identity was later formally
proved by Sherman [4-6], followed later on by another proof by Burgoyne [7].
A somewhat similar treatment to the combinatorial formulation
of Kac, Ward and Feynman can be found in refs. 
[12-14].
An important variant of the combinatorial formulation using the so called
Pffafians was developed by Green
and Hurst [10]. 

The bibliography on the Ising model is vast and to give a full list
of references is virtually impossible. 
A nice introduction to the model though is the paper
by B. Cipra given in ref. [17].
Old surveys but still useful
 on the distinct formulations
of the Ising
model in two dimensions and its history 
can be found in refs. [10-11, 15-16] 
together with  full lists of original
references.

The objective of the present paper is to review in a selfcontained
way the calculation
of the Onsager's formula for the two dimensional free field Ising model
in the combinatorial formulation of Kac, Ward and Feynman.
Our presentation follows chapter 5, section 5.4,
of  Feynman's book [9]
and the paper by Burgoyne [7] although we have tried to be more
careful with the mathematics involved than these references are.
    
The paper is organized as follows.
In section 2, the Ising model is defined. In section 3 and through
its various subsections the combinatorial
formulation of Kac, Ward and Feynman of the partition function is given. 
In section 4,
Onsager's formula
for the free energy per site in the thermodynamic limit is 
computed.

\section{ Definition of the model.} 

The model is defined
on a finite planar  square lattice  $\Lambda$
which mimic a regular arranjement of
atoms in two  dimensions. Suppose the lattice is embedded in the plane with
sites having coordinates in ${\bf Z \times Z}$.
To each  site $i$ of $\Lambda$ it is assigned two possible states also called ``spins'' and
denoted by
$\sigma_{i}$, where $\sigma_{i}=+1$ or $\sigma_i=-1$. 
The interaction energy between two particles located
at the $i$-th and $j$-th sites and in the states $\sigma_{i}$
and $\sigma_{j}$, respectively, is postulated to be
$$
E_{ij}= \left\{ \begin{array}{ll}
               -J \sigma_{i} \sigma_{j}  &  \mbox{if $i, j$ are n.n.}\\
                0 &  \mbox{  otherwise}
\end{array} \right .
$$
$$
\eqno(2.1)
$$
where ``n.n'' stands for nearest neighbors, hence, in the Ising model it is 
 assumed that the energy depends only on short range 
interactions.
 The energy is $-J$ if the nearest
neighbors are in the same state and $+J$ if the states 
are distinct. The constant $J$ which can be positive or negative
is a parameter for the model.

\noindent Suppose $\Lambda$ has $N^{2}$ sites. Then,
there are $2^{N^{2}}$
distinct configurations of the spins and, therefore, $2^{N^{2}}$ configurations
$\sigma=(\sigma_{1},...,\sigma_{N^{2}})$ of the system. Call $S=\{ \sigma \}$ 
the set of possible configurations of the system.
The energy of each configuration $\sigma \in S$ is given by
$$ 
E_{\sigma}=-J\sum _{n.n. \in \sigma}\sigma _i \sigma _j
\eqno(2.2)
$$
\noindent Suppose as well the system is 
at equilibrium  temperature given by $T$. According to statistical mechanics,
the probability $p_{\sigma}$ to find the system in the configuration
 $\sigma$ is
$$
p_\sigma=\frac{1}{Z(\beta)}e^{-\beta E_\sigma}
\eqno(2.3)
$$
where $\beta=\frac{1}{k_B T}$, ${k_B}$
is Boltzmann constant, and 
$$
Z(\beta)=\sum_\sigma e^{-\beta E_\sigma}.
\eqno(2.4)
$$
is the so called partition function of the model. This simple looking
function is simple to compute exactly only in one dimension, difficult
but possible to compute exactly in dimension two.  In three dimensions
nobody knows how to do it.

The exact knowledge of $Z(\beta)$ allows one to obtain
information about the global behaviour of the system. 
Important quantities that are relevant to understand the
physics of the system are all defined in terms of $ln Z$ or its
derivatives.
For instance, 
the free energy per lattice site $f$ in the thermodynamic limit is defined as 
$$
f=-k_{B}T \lim_{N \rightarrow \infty }\frac{\log
Z}{N^{2}}.
\eqno(2.5)
$$
A basic problem is to find a closed form, analytic expression for $f$.
Phase transitions will appear as singularities in $f$ or in one of its 
derivatives.

\section{The combinatorial formulation}

In the combinatorial formulation the partition function is
expressed
as a sum over special subsets of the lattice
$\Lambda$ called admissible graphs.    
Next, using a relation first conjectured by R. Feynman
the resulting expression is converted
into a product over  paths.
 The final  step towards the Onsager's formula to be accomplished
in section 4 consists
in deriving an integral representation for this product.

\subsection{ The partition function as a sum over graphs.}

Let's rewrite the partition function (2.4) as 
$$
Z_{N}(K)=\sum_{\sigma_{1}=\pm 1}\cdots \sum_{\sigma_{N}
=\pm 1}\prod_{n.n.}e^{K\sigma_i\sigma_j}
\eqno(3.1)
$$
with $K=+\frac {J}{k_{B}T}$.
Noting that $\sigma_{i}\sigma_{j}=\pm 1$, it follows that
$$
e^{K\sigma_i\sigma_j} = e^{\pm K}
=\cosh{K} \pm  \sinh{K} 
\eqno(3.2)
$$
and
$$
\prod_{n.n.} e^{K\sigma_{i}\sigma_{j}}
=(1-u^{2})^{-\frac{x}{2}} \prod_{n.n.}(1+\sigma_{i}\sigma_{j}u).
\eqno(3.3)
$$
where $u=\tanh K$ and $x=2N(N-1)$ is the number of bonds in $\Lambda$.
Notice that $\mid u \mid < 1$, for any $K$.

\noindent{\bf Definition 3.1.} An admissible graph is a connected or 
disconnected 
subset of $\Lambda$ whose sites have even valence.

\noindent{\bf Definition 3.2.} Given an admissible graph $G$, define
$$
I_{G}(u)=\prod_{i\in G}u=u^L
\eqno(3.4)
$$
where the product is over the bonds $i$ of $G$.

\noindent{\bf Theorem 3.1.} Call ${\cal A}$ the set of all 
admissible graphs $G$ of $\Lambda$.
Then,
$$
Z_{N}(u)=2^{N^{2}} (1-u^{2})^{-N(N-1)}\left (1+ \sum_{G \in {\cal A}}  I_{G}(u)\right )
\eqno(3.5)
$$

\noindent{\bf Proof:}
To each pair $i,j$ of nearest neighbors
of $\Lambda$ there correspond a term $u\sigma_i\sigma_j$ and a bond.
Since the number of pairs $i,j$ of n.n.
coincide with the number $x=2N(N-1)$ of bonds of $\Lambda$  the 
product on the RHS of (3.3) is a
polinomial of degree $x$, that is, 
$$
\prod_{n.n.}(1+u\sigma_i\sigma_j)
=1+\sum _{p=1}^{x}u^p \sum_{n.n.}(\sigma_{i_{1}} \sigma_{i_{2}})\cdots
(\sigma_{i_{2p-1}}\sigma_{i_{2p}})
\eqno(3.6)
$$
The second summation is over all possible products of $p$ pairs 
$(\sigma_{i}\sigma_{j})$ of n.n. of $\Lambda$ where a pair is not to
occur twice in the same product. To each pair 
$(\sigma_{i}\sigma_{j})$ there is associated a 
bond connecting the neighbors $i$ and $j$ so to each
product of $p$ pairs correspond a graph (connected or disconnected). 
So, the second summation is over
all graphs with $p$ bonds. 
The graphs may have sites with valence 
 $1,2,3$ or $4$. The summations over the spins $\sigma_{i}$'s
 eliminate graphs having
sites with odd valence because
$\sum \sigma_{i} =0 $ and $\sum \sigma_{i}^3=0$. 
The graphs left are those whose
sites have valence 2 or 4, thus admissible. 
If $V_{G}$ is the number of sites in a admissible graph $G$
then there is
a factor $2^{V_{G}}$ associated to it
because each site of $G$ contributes a factor $2$
coming from  $\sum \sigma_{i}^{2} =2 $ and $\sum \sigma_{i}^{4} =2 $. 
The sum over $\sigma$ includes all the $\sigma_{i}$ and not only those
with sites $i$ in $G$. The summation over the sites not in $G$ will give
a factor $2^{V-V_{G}}$ where $V=N^{2}$ is the number of sites in
the lattice, hence, in the end one gets the factor $2^{V}$.

\subsection{ The partition function as a product over paths.}

Let's orient and  number the bonds of $\Lambda$ with distinct
positive integers
$i$ and call $\Lambda$ with this indexation a colored lattice.

\noindent{\bf Definition 3.3.} 
A path $p$ over $\Lambda$ is an ordered sequence of bonds each
starting at the site where the previous one ended.
The last bond ending at the site from which the
first one started. Thus, $p$ is closed. The path
is subjected to the constraint that it never
goes backwards over the previous bond. 
A path  $p$ is given by a word, that is,
an ordered sequence of symbols
$D_{i}$ where $i$ distiguishes the
bonds of $\Lambda$. A path $p$ is then of the form
$$
p =
D_{j_{1}}^{e_{1}}D_{j_{2}}^{e_{2}}
...
D_{j_{l}}^{e_{l}}
\eqno(3.7)
$$
for some $l$ and
where $e_{i} = +1 (-1)$ if the path traverses
bond $j_{i}$ following the direction (opposite direction)
assigned to it.
Because a 
path is closed it is defined to within its circular order
so that
$$
D_{j_{1}}^{e_{1}}D_{j_{2}}^{e_{2}}
...
D_{j_{l}}^{e_{l}}
\equiv
D_{j_{2}}^{e_{2}}
...
D_{j_{l}}^{e_{l}}D_{j_{1}}^{e_{1}}
\equiv ... \equiv 
D_{j_{l}}^{e_{l}}
D_{j_{1}}^{e_{1}}...D_{j_{l-1}}^{e_{l-1}}
\eqno(3.8)
$$
The inversion $p^{-1}$ of $p$ is given by
$$
D_{j_{l}}^{-e_{l}}D_{j_{l-1}}^{-e_{l-1}}
...
D_{j_{1}}^{-e_{1}}
\eqno(3.9)
$$
We take $p$ and $p^{-1}$ to be equivalent. Given $p$, denote by
$[p]$ the set of all 
paths equivalent to $p$, that is, its circular permutations
and their inversions.
 
\noindent{\bf Definition 3.4.} 
A periodic path is one which
has the word representation
$$
(D_{j_{1}}^{e_{1}}
...
D_{j_{l}}^{e_{l}})^{w}
\eqno(3.10)
$$
for some $l$ and $w \geq 2$
and where the subword in between brackets is nonperiodic.

\noindent{\bf Definition 3.5.} A path $p$ has assigned to it a sign given by
$$
s(p) = (-1)^{1+t}
\eqno(3.11)
$$
where $t$ is the number of $2 \pi$-angles
turned by a tangent vector while 
traversing $p$. A positive (negative) angle is assigned to a 
counterclockwise (clockwise)
rotation.

\noindent {\bf Example 1.} See Figure 1a). 
A tangent vector starting at point $e$ 
and traversing the path shown
in Figure 1a) turns once a total angle given by $4. \frac{\pi}{2}= 2 \pi$
after its return to $e$ so in this
case $t=1$ and $s(p)=+1$. For the path in Figure 1b), the total angle turned is
$3 \frac{\pi}{2}- 3 \frac{\pi}{2}=0$  so $t=0$ and $s(p)=-1$.

\begin{picture}(300,165)(-160,-20)
  \put(-80,50){\line(1,0){35}}
  \put(-80,50){\line(0,1){35}}
  \put(-80,85){\line(1,0){35}}
  \put(-45,50){\line(0,1){35}}
  \put(-52.5,50){\circle*{3}}
  \put(-52.5,40){e}
  \put(-62,50){\vector(1,0){1}}
  \put(-80,68){\vector(0,-1){1}}
  \put(-45,68){\vector(0,1){1}}
  \put(-62,85){\vector(-1,0){1}}
  \put(-68,20){(a)}

  \put(70,50){\line(1,0){35}}
  \put(70,50){\line(0,1){35}}
  \put(70,85){\line(1,0){35}}
  \put(105,50){\line(0,1){35}}
  \put(97.5,50){\circle*{3}}
  \put(97.5,40){e}
  \put(88,50){\vector(1,0){1}}
  \put(70,68){\vector(0,-1){1}}
  \put(105,68){\vector(0,1){1}}
  \put(88,85){\vector(-1,0){1}}

  \put(105,120){\line(1,0){35}}
  \put(105,120){\line(0,-1){35}}
  \put(105,85){\line(1,0){35}}
  \put(140,120){\line(0,-1){35}}
  \put(122.5,120){\vector(1,0){1}}
  \put(122.5,85){\vector(-1,0){1}}
  \put(140,102.5){\vector(0,-1){1}}

  \put(82,20){(b)}
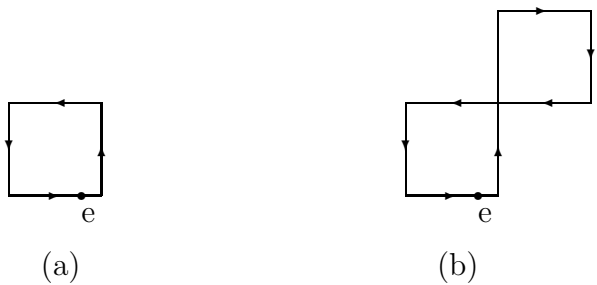
  \put(-150,-10){Figure 1: Examples of paths with (a) $s(p)=+1$ and
              (b) $s(p)=-1$. }
\end{picture}

\noindent {\bf Remark.} In section 4, instead of assigning 
an angle $\pm \pi/2$ to a turn
we will count the contribution to the sign by 
assigning
$\alpha = e^{i \pi/4}$ and $\bar{\alpha}=e^{-i \pi/4} $ to each 
counterclockwise and clockwise turn, respectively, and then in the 
end multiplying
the result by $-1$. In the example above, one gets in this manner
$(e^{i \pi/4})^{4}= -1$ and $(e^{i \pi/4})^{3}. (e^{-i \pi/4})^{3}=+1$.
Multiplying both results by $-1$, one recovers the correct sign for each path.

\noindent {\bf The sign of periodic paths.} Suppose the sign of the 
nonperiodic path in between brackets in (3.10) is $(-1)^{1+t_{s}}$.
Then, the sign of the periodic path with period $w$ is
$(-1)^{1 +wt_{s}}$. Hence, the sign of a periodic path is $-1$ if
its period is an even number and the sign equals the sign of the nonperiodic
subpath if the period $w$ is an odd number.

\noindent{\bf Definition 3.6.} To each path $p$ it is assigned the 
function
 $I_{p}(u)$ given by
$$
I_{p}(u)=u^{l}
\eqno(3.12)
$$
where $l=m_{1}+...+m_{k}$, for some $k$,
 is the length of $p$, $m_{i}$ being the
number of times bond $i$ is covered
by $p$, and the function
$W_{p}(u)$, ``the amplitude of $p$'', defined as follows:
$$
W_{p}(u)=s(p) I_{p}(u)
\eqno(3.13)
$$
\noindent{\bf Theorem 3.2.} The functions $I_{G}(u)$ and $W_{p}(u)$,
$|u| < 1$, 
defined above
satisfy the following relation:
$$
1+\sum_{G \in {\cal A}}I_{G}(u) =\prod _{[p] }[1+W_{p}(u)]
\eqno(3.14)
$$
The product is over all inequivalent classes $[p]$
of closed nonperiodic paths.
The summation
is over all admissible graphs of the finite $N \times N$
planar square lattice $\Lambda$.

Relation (3.14) is a simpler version suitable for the Ising model of
a more general relation investigated by Sherman and Burgoyne in refs. [4-7]. 
The
difference is that they assign to the bonds $i$ of the lattice
distinct parameters $d_{i}$, hence, in this case 
the functions $I_{G}$ and $W$ are given
in terms of these parameters.
In the Ising model context under consideration
 these are all equal to $u$ and $|u|<1$. 

According to references [4,7,10,11], relation (3.14)
first appeared as a conjecture in lecture notes by
Feynman ( ref. [9], published only in 1972 and already mentioning ref. [4]). 
The first proof of it was achieved by Sherman in refs. [4,6] followed
by another one later on by Burgoyne in ref. [7]. The simplest nontrivial
case of the general relation is investigated in ref. [8].

Below Burgoyne's proof is essencially reproduced
for the case $|u| < 1$.

\noindent {\bf Proof:} Expand the product over the 
distinct classes of nonperiodic
paths $[p]$ as $1$ (one) plus an infinite sum
of terms of the form 
$$
W_{p_{1}}(u)W_{p_{2}}(u)...W_{p_{k}}(u)= s \prod_{i} u^{r_{i}}
\eqno(3.15)
$$
for some $k$ where $p_{1},...,p_{k}$
is a set of nonperiodic paths over $\Lambda$.
The product on the r.h.s of (3.15) is over the bonds $i$ traversed
by $p_{1},p_{2},...p_{k}$,
and $r_{i}$ says how many times.  
If 
$p_{1}, p_{2},...p_{k}$ traverse bond $i$, say, $m_{1}(i),...,m_{k}(i)$
times, $m_{j} \geq 0$, respectively, then $r_{i}=\sum_{i=1}^{k}m_{j}(i)$.
The  sign $s$ is the product of the signs
of $p_{1}, p_{2},...,p_{k}$.

Let's prove, first, that those terms having
$r_{i}=1$, $\forall i$, add up to $\sum I_{G}(u)$. 
Consider one of these terms with associated paths $p_{1}, p_{2}, ..., p_{k}$.
Each bond in the set of bonds traversed by 
$p_{1}, p_{2},..., p_{k}$ 
is traversed only once by one of these paths. Thus, the only
possible intersection if any between any two of these paths in
this case can occur only at a site
of valence 4 and they cross each other like in Fig. (2.a) . 
Otherwise, they are disjoint. 
Thus, the set of bonds traversed by paths $p_{1},...,p_{k}$
constitute a graph whose vertices have valence $2$ or $4$. This
is an admissible graph. Therefore, to each term of the form of (3.15)
with $r_{i}=1$, $\forall i$, one can associate an admissible graph.
This graph can be disconnected. This happen if
the set of paths can be split into subsets completely disjoint
which generate admissible graphs without any bonds and vertices in common.

Now, given an admissible graph $G$ one can in general associate
more than one term of the form of $(3.15)$ with $r_{i}=1$, each associated
with a distinct
set of paths. Let's see how this follows. The sites of an admissible graph
have valence $2$ or $4$. When a path strikes a site of valence $4$ 
it has only $3$ possible directions
to follow. See Figures 2a, 2b, 2c. (The case in Fig. 2d is forbidden.)
Then, any two terms associated to a given admissible graph
$G$ will differ only in the types of crossings at
the sites of $G$. Since there are $3$ types of crossing per valence 4
site, 
the number of possible terms associated to $G$ is $3^{V}$ where $V$ 
is the number of sites of $G$ with valence $4$. 

A term has a sign which comes out
from the contribution of the signs of the paths associated to that term. Let's see how
the sign of a term comes out.
A term with $t_{1}$ crossings of type $1$ (Fig. 2a.) has a
sign which can be expressed as 
$(-1)^{t_{1}}$ where $t_{1}$ 
includes selfcrossings of single paths plus crossings between different paths. 
Indeed, since distinct closed paths always intersect in a 
even number of crossings then $(-1)^{t_{1}}$ will give 
the correct sign of the term which
is the product of the signs of the individual paths.
Let's associate to the crossings of type $j=2,3$ the sign
$(+1)^{t_{2}}(+1)^{t_{3}}$
so that a term with $t_{j}$ crossings of type $j=1,2,3$
has a sign given by $(-1)^{t_{1}}(+1)^{t_{2}}(+1)^{t_{3}}$.

There are $V!$ ways of distributing $V=t_{1}+t_{2}+t_{3}$ crossings among the sites of $G$
but since there are $t_{j}$ crossings of the type $j$, $j=1,2,3$,
 one has to divide
$V!$ by $t_{1}!t_{2}!t_{3}!$ so that the number of distinct terms with
$t_{j}$ crossings of type $j$ is
$$
\frac{V!}{t_{1}!t_{2}!t_{3}!}
\eqno(3.16)
$$
These terms have the same factor $I(G)=u^{L}$
where $L$ is the number of bonds of $G$.
Summing all these terms arising from a given $G$ and 
summing over all admissible graphs $G$ of $\Lambda$
 the result is 
$$
\left(
\sum_{G} \sum_{\{t\}}\frac{V!}{t_{1}!t_{2}!t_{3}!} 
(-1)^{t_{1}}(+1)^{t_{2}}(+1)^{t_{3}} \right)I(G)
\eqno(3.17)
$$
where $\sum_{t}$ means summation over all $t_{1}, t_{2}, t_{3}$ such that 
$t_{1}+t_{2}+t_{3}=V$. Using the multinomial theorem  the summation over
$\{t\}$ gives $(-1+1+1)^{V}$ and one gets the result $\sum _{G}I(G)$.

If $G$ is  disconnected with $l$ components $G_{i}$, $i=1,2,...,l$,
each of them
with $t_{j}$,
$j=1,2,...,l$, sites of valence $4$ and $\sum t_{j}=V$, 
then applying the
previous argument to each component  will give $I(G_{1})I(G_{2})...I(G_{l})=
I(G)$. 

In view of the above result, 
the theorem could be equivalently stated by saying that the sum of 
terms with $r_{i} > 1$ for at least one of the $i$ converges to zero. 
Let's prove this. 

Let ${\cal G}$ be the set of all colored connected or disconnected subgraphs
$g$ of the colored lattice without valence 1 sites and 
such that if $g$ is connected then $g$ is 
not a poligon, that is, a graph having valence 2 sites only.
A disconnected graph is allowed to have some but not all of the components as poligons.
The reason for excluding graphs which are poligons
or having all
components which are 
is that closed paths with repeated bonds over them
are necessarily periodic and these are forbidden.
The coloring of $g$ is that inherited from 
the colored lattice. 

Given
$g \in {\cal G}$, call $i_{1},...,i_{l(g)}$ the bonds of $g$.
A term  $w_{g}$ associated to  $g$ is  of the form
$$
w_{g}=W_{p_{1}}W_{p_{2}}...W_{p_{k}}= (sign \hspace{1mm} w_{g}) |w_{g}|
\eqno(3.18)
$$
for some $k$ and set of paths $p_{1},...,p_{k}$, which traverse
the bonds of $g$ only, where
$$
|w_{g}|= 
\prod_{j=1}^{l(g)}u^{r_{i_{j}}}
\eqno(3.19)
$$
and $r_{i_{j}}$ is the number of times bond $i_{j}$  
is traversed by $p_{1},...,p_{k}$, that is,
If $p_{1},..., p_{k}$ traverse the $i$-th bond
 $m_{1}(i),...,m_{k}(i)$ times, $m \geq 0$, respectively, 
then 
$$
r_{i_{j}}=\sum_{a=1}^{k} m_{a}(i_{j})
\eqno(3.20)
$$
Some but not all of the $m$'s can be zero so that $r_{ij} \geq 1$ with at
least one $r_{ij} > 1$. 

Let's consider the set of all terms with the same effective set of bonds 
$\{i\}$ traversed, hence, the terms associated to a given $g$.
Within this set it's possible in general 
to find terms with the same powers $\{r\}$ and the same $|w_{g}|$
although having  distinct associated paths and possibly with  different
effective sign.

Let's group together those terms which cover
the same  bonds of $g$ the same number of times. 
Denote by ${\cal W}_{g,N}(r)$ the set of terms  $w_{g}$
with the same  powers $\{r_{i_{j}}\}$ and  such that
  $\sum_{j=1}^{l(g)} r_{i_{j}}=N$, for fixed $N$.
The summation over all terms with repeated lines can now be expressed as
$$
\sum_{g\in \mathcal{G}}\quad
\sum_{\{N\}_{g}} \quad \sum_{r(N)} 
\quad \sum_{w_{g}\in {\cal W}_{g,N}(r)} w_{g}
\eqno(3.21)
$$
where $\sum_{g \in {\cal G}}$ means summation over all elements in 
${\cal G}$; $\sum_{ \{N\}_{g}} $ means summation over all
positive integers $N$ compatible to the given graph $g$ and such that
$N \geq l(g)+1$;
$\sum_{r(N)}$ means summation over a set of positive
integers $r_{1},
...,r_{l}$ such that $r_{1}+...+r_{l}=N$ and which are also compatible
to $g$; and, finally, $\sum_{w_{g}}$ means summation over all terms
$w_{g} \in {\cal W}_{g,N}(r)$.

Now the following remarks  come to order.
In the second summation,
the case $N=l$ is excluded for it implies that $r_{i}=1$ and in this case
there can be no repeated bonds. The case $N < l$ corresponds to 
another
element $g' \in {\cal G}$. The equality depends on the graph $g$.
For instance,
take the graph shown in Fig. 1b where $l+1=9$. No nonperiodic
closed path with repeated bonds can have length $N=9$ because $l(g)=8$.
The length $N$ can only be even and its minimum is $N=12$. Hence, for 
this particular graph $g$ the
summation is over all even numbers greater or equal to 12.
In any case, the set $\{N\}_{g}$ has always infinite elements.
Given $g$ and  $N \in \{N\}_{g}$,
not all partitions of $N$ are allowed in the third sum. For instance,
given the graph in Fig. 1b
and $N=12$, the partition with $r_{i_{k}}=1$, $\forall k \neq 1$, and 
$r_{i}=5$ can not be associated to any allowed path.
So, the set of integers $\{N\}$ and partitions of $N$ must be suitable to
each $g$.

Given $g$, let's consider now the partial sums
$$
s_{n}=\sum_{\{N | N \leq n\}_{g}} \quad \sum_{r(N)} 
\quad \sum_{w_{g}\in {\cal W}_{g,N}(r)} w_{g}
\eqno(3.22)
$$
The goal is to show that in the limit $n \rightarrow \infty$, $s_{n}$
goes to zero. In ref. [7] its proved that $s_{n}=0$. The argument of the
proof goes as follows.

Since the bonds of $g$ are covered the same number of times by all elements
in the group,
choose a bond of $g$, say $b$, which is traversed $\overline{r} > 1$ 
times by all elements in the group
${\cal W}_{g,N}$. This choice has to be done for each partition $r(N)$.
Denote by $P$ the set of paths associated to $w_{g}$. 
Then, $P=P' \bigcup P''$ where $P'$ 
is the set of those paths which traverse
bond $b$ whereas $P''$ is the set of those paths  which do not
traverse $b$.

Given a path $p \in P'$, let $p_{c}$ be the path segment obtained from $p$ 
upon removal of $b$. Given $P' = \{p,p'...\}$ define
$$
P_{c}'=\{p_{c},p_{c}',...\}
\eqno(3.23)
$$
This set has exactly $\overline{r}$ path segments.
Collect under a same subgroup $S$ the elements $w_{g}\in {\cal W}_{g,N}$ 
having the 
property that line $b$ is covered exactly $\overline{r}$ times by all elements
in $S$ and
they all have the same subset $P_{c}'$ with
$\overline{r}$ path segments and the same subset $P''$.
The set ${\cal W}_{g,N}$ is the union of such subsets, that is,
$$
 \sum_{w_{g}\in {\cal W}_{g,N}(r)} w_{g} =
 \sum_{S\subset
{\cal W}_{g,N}(r)} \quad \sum_{w_{g} \in S}s(w) |w_{g}|
\eqno(3.24)
$$
where $s(w)$ is the sign of $w_{g}$ and $|w_{g}|= u^{N}$. Recall that
$|u|<1$ so that $ |w_{g}| < 1$.

The elements inside any given $S$ cancel
each other. 
Denote by $q$ and $e$
the elements of $S$ that are in $P'$ and $P''$, respectively. 
Suppose
that the segments $q_{1},...,q_{ \overline{r}}$ are all distinct. 
(For the case with repeated segments, see [7]).
The terms in $S$ 
are precisely 
those which can be obtained by joining the ends of the segments and
this can be done in exactly $\overline{r} !$ ways.
This gives the possible terms $w_{g}$ in the subgroup. From the 
properties of the permutation group half of $N!$ permutations are odd
and half are even and so the signs of half of the terms are positive
and half are negative, hence, a cancellation takes place.


Using (3.14), the partition function of the two dimensional
Ising model can now be expressed as a product over paths as follows:
$$
Z_{N}(u)=
2^{N^{2}}(1- u^{2})^{-N(N-1)} \prod_{[p]} \left[ 1+W_{p}(u) \right]
\eqno(3.25)
$$
The next step  consists
in expressing the product over $[p]$ as an integral. This will be 
achieved in the next section.

\section{ Paths amplitudes and Onsager's formula}

Consider all paths that start at a fixed site
$P_{1}$ 
which we take as the origin with coordinates $(0,0)$ and end at the site
$P_{n+1}$ with coordinates
$(x,y)$ in $n$ steps. 
Starting at $(0,0)$ and whenever a site is reached
 there are
four possible directions which a path can take (see Figure 2 and the
Remark below).
The path {\bf a)} continues forward in the same direction of the previous step;
{\bf b)} it turns left $90^0$ relative to the previous step;
{\bf c)} it turns right $90^0$ relative to the previous step;
{\bf d)} it turns $180^0$.
To each one of this possibilities it is assigned an ``amplitude''
which is given by:
{\bf A)} $u$ for the case {\bf a)};
{\bf B)} $u\alpha$ for the case {\bf b)};
{\bf C)} $u \bar{\alpha}$ for the case {\bf c)} and
{\bf D)} $0$ (zero) for the case {\bf d)},
 where $u=tgh K$ and $\alpha= e^{i\pi/4}$ is  the contribution to the
 sign of $p$ each time it turns left (counterclockwisely)
relative to the previous step and $\bar{\alpha}$, when it turns right
(clockwisely). See the Remark after Example 1, sec. 3.2.

\begin{picture}(300,135)(-100,-25)
  \put(-80,55){\line(1,0){50}}	
  \put(-55,30){\line(0,1){50}}	
  \put(-55,15){(a)}
  \put(-50,55){\vector(1,0){10}}
  \put(-75,55){\vector(1,0){10}}

  \put(10,55){\line(1,0){50}}	
  \put(35,30){\line(0,1){50}}	
  \put(35,15){(b)} 
  \put(15,55){\vector(1,0){10}}
  \put(35,60){\vector(0,1){10}}

  \put(100,55){\line(1,0){50}}	
  \put(125,30){\line(0,1){50}}
  \put(105,55){\vector(1,0){10}}
  \put(125,50){\vector(0,-1){10}}
  \put(125,15){(c)}	

  \put(190,55){\line(1,0){50}}	
  \put(215,30){\line(0,1){50}}	
  \put(192,55){\vector(1,0){10}}
  \put(216,55){\vector(-1,0){10}}
  \put(215,15){(d)}
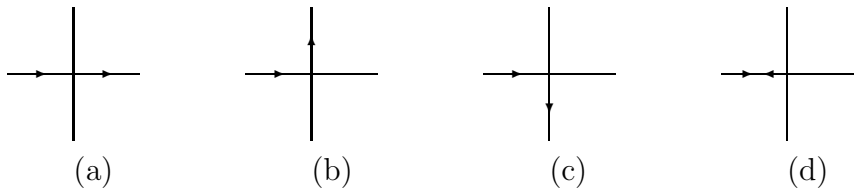
\put(-80,-10){Figure 2: Directions which a path can take at a valence 4
site. }
\end{picture}

\noindent {\bf Remark.} The lattice being finite it has a border
so that when a path strikes a site on the border it may have there only
two or three
possible directions to follow. 
In the spirit of refs. [7,9] we shall neglect the border and derive the relevant 
formulas as if there was no border at all with the justification
that in the limit 
$N \rightarrow \infty$
which we shall take in the end of the calculation border effects 
dissapear. Of course, another approach would be
to do everything on a toroidal lattice. In this case, however, relation (3.14)
must be replaced by another more involved identity apropriate for the toroidal
lattice ( given in refs. [4, 10] ). We shall restrict
the presentation to the planar case only.

Call
$U_{n}(x,y)$ the amplitude of arrival at $(x,y)$ moving upward in the $n$-th
step,
$D_{n}(x,y)$ the amplitude of arrival at $(x,y)$ moving downwards in the
$n$-th step,
$L_{n}(x,y)$ the amplitude of arrival at $(x,y)$ moving from the left in the
$n$-th step, and
$R_{n}(x,y)$ the amplitude of arrival at $(x,y)$ moving from right in the
$n$-th step.

\noindent If the path arrives at $(x,y)$ moving upward in the n-th step then
$$
U_{n}(x,y)=uU_{n-1}(x,y-1)+0D_{n-1}(x,y-1)+u\alpha L_{n-1}(x,y-1)
+u{\bar{\alpha}}R_{n-1}(x,y-1)
$$
$$ 
\eqno(4.1)
$$
where $U_{n-1}$, $D_{n-1}$, $L_{n-1}$ and $R_{n-1}$ are the
amplitudes associated to the four possibilities to reach
site $(x,y-1)$. Relation (4.1) can be understood as follows.
If $(x,y-1)$ is reached going up a bond in the $(n-1)$-th step,
there the amplitude is $U_{n-1}(x,y-1)$ so in 
the   $n$-th step as the path follows the same direction of the 
previous step, by the rules $a)$ and $A)$ above,
a factor $u$ is multiplied to the amplitude $U_{n-1}(x,y-1)$. See Figure 3.

\begin{picture}(300,190)(-140,-40)
  \put(35,30){\line(0,1){100}}	
  \put(35,105){\vector(0,1){1}}
  \put(40,130){(x,y)}
  \put(35,130){\circle*{3}}
  \put(35,80){\circle*{3}}
  \put(35,55){\vector(0,1){1}}
  \put(40,80){(x,y-1)}
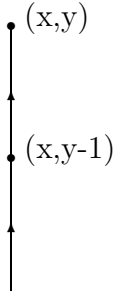
\put(-160,-10){Figure 3: $p$ goes up to $(x,y-1)$ and  $(x,y)$
in the $(n-1)$-th
 and $n$-th steps. }
\end{picture}

\noindent If the site $(x,y-1)$ is reached from the left
in the $(n-1)$-th step (Figure 4),
the path has to  make a
counterclockwise rotation to go to $(x,y)$ in the $n$-th step. 
By the rules $b)$ and $B)$
a factor $u \alpha$ should then be multiplied to the amplitude
$L_{n-1}(x,y-1)$.

\begin{picture}(300,135)(-150,-40)
  \put(35,30){\line(0,1){50}}	
  \put(-15,30){\line(1,0){50}}
  \put(35,55){\vector(0,1){1}}
  \put(10,30){\vector(1,0){1}}
  \put(40,80){(x,y)}
  \put(35,80){\circle*{3}}
  \put(35,30){\circle*{3}}
  \put(40,30){(x,y-1)}
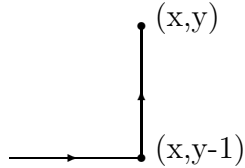
\put(-170,-10){Figure 4. $p$ turns counterclockwise at $(x,y-1)$ to go
to $(x,y)$ in the n-th step. }
\end{picture}

\noindent The case that the path goes down to $(x,y-1)$ in the
$(n-1)$-th step and goes up to $(x,y)$ 
in the $n$-th step
corresponds to a
$180^{0}$  rotation. By rules  $d)$ and $D)$ the amplitude should be
 $0.D_{n-1}(x,y-1)$. If the site
$(x,y-1)$ is reached from the right the path has 
to make a clockwise rotation to go to $(x,y)$ (Figure 5). 
By the rules  $c)$ and $C)$  a factor
$u\bar{\alpha}$ should then be multiplied to
the  amplitude $R_{n-1}(x,y-1)$.

\begin{picture}(300,135)(-120,-40)
  \put(35,30){\line(0,1){50}}	
  \put(35,30){\line(1,0){50}}
  \put(35,55){\vector(0,1){1}}
  \put(55,30){\vector(-1,0){1}}
  \put(40,80){(x,y)}
  \put(35,80){\circle*{3}}
  \put(35,30){\circle*{3}}
  \put(23,13){(x,y-1)}
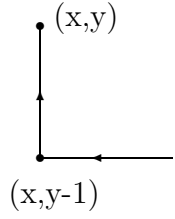
\put(-130,-10){Figure 5.   $p$ turns clockwise at $(x,y-1)$ to go
to $(x,y)$ in the n-th step.    }
\end{picture}

\noindent Analogously,
if a path arrives at $(x,y)$ in the $n$-th step going down
the amplitude is given by the relation
$$
D_{n}(x,y)=0U_{n-1}(x,y+1)+uD_{n-1}(x,y+1)+u{\bar{\alpha}}L_{n-1}(x,y+1)
+u\alpha R_{n-1}(x,y+1) 
$$
$$
\eqno(4.2)
$$
If it arrives at $(x,y)$ coming from
the left then the amplitude is given by
$$
L_{n}(x,y)=u\bar{\alpha}U_{n-1}(x-1,y)+u\alpha D_{n-1}(x-1,y)+uL_{n-1}(x-1,y)
+0R_{n-1}(x-1,y)
$$
$$ 
\eqno(4.3)
$$
At last,
if it arrives at  $(x,y)$ coming from the right 
the amplitude is
$$
R_{n}(x,y)=u\alpha U_{n-1}(x+1,y)+u\bar{\alpha}D_{n-1}(x+1,y)+0L_{n-1}(x+1,y)
+uR_{n-1}(x+1,y)
$$
$$ 
\eqno(4.4)
$$
Of course to compute an amplitude using the above recursion relations 
it is needed the amplitude in the zero-th step. We shall follow the 
 convention
of reference [10], namely,
that in the zero-th step a path arrives at the origin moving upward
so that $U_{0}(x,y)=\delta_{x,0} \delta_{y,0}$ and 
$D_{0}=R_{0}=L_{0}=0$. The amplitude to
arrive in zero steps is one if the path arrives going upward at the origin and zero for any
other point or any other direction of arrival.

\noindent{\bf Example 2.} See Figure 6.
Let's compute the amplitude of arrival at site (2,1) in 3 steps moving upward
in the third step. Only one path is possible in this case.
Using the recursion (4.1),
$$
U_{3}(2,1)=uU_{2}(2,0)+0D_{2}(2,0)+u\alpha L_{2}(2,0)
+u{\bar{\alpha}}R_{2}(2,0)
\eqno(4.5)
$$

\begin{picture}(250,155)(-300,-50)
  \put(-95,30){\line(0,1){35}}
  \put(-95,30){\line(-1,0){70}}
  \put(-130,30){\circle*{3}}
  \put(-105,75){(2,1)}
  \put(-95,65){\circle*{3}}
  \put(-95,30){\circle*{3}}
  \put(-165,30){\circle*{3}}
  \put(-180,15){(0,0)}
  \put(-140,15){(1,0)}
  \put(-105,15){(2,0)}

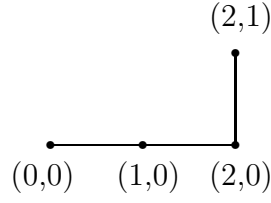
\put(-230,-40){Figure 6: The path  in  Ex. 2.}

\end{picture}

\noindent In the second step, the path moves to site (2,0) coming from the left
so $U_{2}=D_{2}=R_{2}=0$ and $U_{3}(2,1)=u \alpha L_{2}(2,0)$. From
(4.3),
$$
L_{2}(2,0)=u\bar{\alpha}U_1(1,0)+u\alpha D_1(1,0)+uL_1(1,0)
+0R_1(1,0)
\eqno(4.6)
$$
with $U_{1}=D_{1}=R_{1}=0$ so that $U_{3}(2,1)=u^{2} \alpha L_{1}(1,0)$
where
$$
L_{1}(1,0)=u\bar{\alpha}U_0(0,0)+u\alpha D_0(0,0)+uL_0(0,0)
+0R_0(0,0)=
u\bar{\alpha}
\eqno(4.7)
$$ 
implying that $U_{3}(2,1)=u^{3}$.

\noindent{\bf Example 3.} Let's now compute the amplitude of arrival  at
(2,1) in 3 steps moving from the left in the third step.
In this case, the possible paths are shown in Figure 7a) and 7b).

\begin{picture}(250,155)(-100,-50)
  
  \put(-30,30){\line(1,0){35}}
  \put(5,30){\line(0,1){35}}
  \put(5,65){\line(1,0){35}}
  \put(-30,30){\circle*{3}}
  \put(5,30){\circle*{3}}
  \put(5,65){\circle*{3}}
  \put(40,65){\circle*{3}}
  \put(5,47.5){\vector(0,1){1}}
  \put(22.5,65){\vector(1,0){1}}
  \put(-10,30){\vector(1,0){1}}
  \put(-43,15){(0,0)}
  \put(-10,15){(1,0)}
  \put(-5,75){(1,1)}
  \put(30,75){(2,1)}
  \put(-5,-10){(a)}

  \put(130,30){\circle*{3}}
  \put(130,30){\line(0,1){35}}
  \put(130,47.5){\vector(0,1){1}}
  \put(120,15){(0,0)}
  \put(120,75){(0,1)}
  \put(130,65){\circle*{3}}
  \put(130,65){\line(1,0){70}}
  \put(165,65){\circle*{3}}
  \put(200,65){\circle*{3}}
  \put(155,75){(1,1)}
  \put(190,75){(2,1)}
  \put(155,-10){(b)}

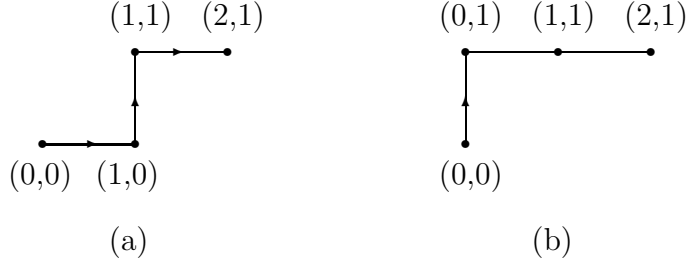
  \put(-20,-40){Figure 7: The paths in Ex. 3.}

\end{picture}

Using relation (4.3), the amplitude is
$$
L_{3}(2,1)=u\bar{\alpha}U_2(1,1)+u\alpha D_2(1,1)+uL_2(1,1)
+0R_2(1,1)
\eqno(4.8)
$$
Using (4.1),
$$
U_{2}(1,1)=uU_{1}(1,0)+0D_{1}(1,0)+u\alpha L_{1}(1,0)
+u{\bar{\alpha}}R_{1}(1,0)
\eqno(4.9)
$$
Since $U_{1}=D_{1}=R_{1}=0$, one finds that 
$U_{2}(1,1)=u \alpha L_{1}(1,0)=u \alpha u \bar{\alpha}=u^{2}$. Using (4.3),
with $D_{1}=L_{1}=R_{1}=0$,
$$
L_{2}(1,1)=u\bar{\alpha}U_1(0,1)+u\alpha D_1(0,1)+uL_1(0,1)
+0R_1(0,1)=u \bar{\alpha} u
\eqno(4.10)
$$
Therefore, $L_{3}(2,1)=2u^{3} \bar{\alpha}$.

\noindent {\bf Definition 4.1. } The partial amplitude of a path $p$ of length
$n$ is given by
$$
\overline{W}_{p}(u)= \prod \alpha \prod \bar{\alpha}  \hspace{1mm} u^{n}
\eqno(4.11)
$$

\noindent {\bf Definition 4.2 .} The amplitude  
$\sum_{p} \overline{W}_{p(n,P_{1})}(u)$ of 
 arrival at $P_{n+1}(x,y)$ from any direction
in $n$ steps is given by
$$
U_{n}(x,y)+D_{n}(x,y)+L_{n}(x,y)+R_{n}(x,y) 
\eqno(4.12)
$$

\noindent {\bf Example 4}. The partial amplitudes for the paths in Figure 6,
7a) and 7b) are $u^{3}$, $\bar{\alpha}u^{3}$ and $\bar{\alpha}u^{3}$,
respectively. The amplitude of arrival at (2,1) from any direction in 3 steps
is, then, $u^{3}+2\bar{\alpha}u^{3}$.

\noindent {\bf Definition 4.3.} Fix $n$ and call $C_{n}(x,y)$ the 
set of all paths of length $n$ starting at $(0,0)$
and arriving at $(x,y)$. Given  $p \in C_{n}$ and $F_{n}
\in {\cal B}(x,y)$ where
$$
{\cal B}(x,y) = \{U_{n}(x,y), D_{n}(x,y), L_{n}(x,y), R_{n}(x,y) \}
\eqno(4.13)
$$
Define the extension of $F_{n}(x,y)$, denoted by the
same symbol, so as to include sites $(x,y)$ which
can be reached only by a number $m > n$ of steps but in this case
set $F_{n}(x,y)=0$.

\noindent {\bf Lemma.} The transform of $F_{n}$, 
the function $\overline{F}_{n}(\epsilon,\eta)$,
$0 \leq \epsilon \leq 2 \pi$ and $0 \leq \eta \leq 2 \pi $,
given by
$$
\overline{F}_{n}(\varepsilon ,\eta )=\sum _{x=-\infty}^{\infty}\sum
_{y=-\infty}^{\infty} F_{n}(x,y)e^{-i\varepsilon x}e^{-i\eta y}
\eqno(4.14)
$$
is well defined and
$$
F_{n}(x,y)=\int_{0}^{2\pi}\int_{0}^{2\pi}
e^{i\varepsilon x}e^{i\eta y}
\overline{F}_n(\varepsilon,\eta)\frac{d\varepsilon
d\eta}{(2\pi)^2} 
\eqno(4.15)
$$
\noindent{\bf Proof:}
$F_{n}(x,y)=0$ for $\mid x \mid > n$
or/and  $\mid y \mid > n$. Then, for fixed $n$ the sums in (4.14) have only
a finite number of terms.

Using (4.14), the transform of $U_{n}(x,y)$
is:
$$
\overline {U}_{n}(\varepsilon ,\eta )=\sum _{x=-\infty}^{\infty}\sum
_{y=-\infty}^{\infty} U_{n}(x,y)e^{-i\varepsilon x}e^{-i\eta y}
\eqno(4.16)
$$
Upon substitution of (4.1), and making the change $\bar{y}=y-1$
it follows that
$$
\overline {U}_{n}(\varepsilon ,\eta )
=ue^{-i\eta}\overline{U}_{n-1}(\varepsilon,\eta)+
0 {{\bar{D}}_{n-1}(\varepsilon,\eta)}+u\alpha e^{-i\eta}\overline{L}_{n-1}
(\varepsilon,\eta)+u{\bar{\alpha}}e^{-i\eta}\overline{R}_{n-1}
(\varepsilon,\eta)
$$
$$
\eqno(4.17)
$$
Similarly, we obtain  $\overline {D}_{n}(\varepsilon ,\eta )$,
$\overline {L}_{n}(\varepsilon ,\eta )$ and  $\overline {R}_{n}(\varepsilon
,\eta )$:
$$
\overline{D}_{n}(\varepsilon,\eta)
=0 \overline{U}_{n-1}(\varepsilon,\eta)
+ue^{i\eta}\overline{D}_{n-1}(\varepsilon,\eta)+ 
u{\bar{\alpha}}e^{i\eta}\overline{L}_{n-1}(\varepsilon,\eta)
+u\alpha e^{i\eta}\overline{R}_{n-1}(\varepsilon,\eta)
$$
$$
\eqno(4.18)
$$
$$
\overline{L}_{n}(\varepsilon,\eta)
=u{\bar{ \alpha} }e^{-i\varepsilon}\overline{U}_{n-1}(\varepsilon,\eta)
+ u\alpha e^{-i\varepsilon}
\overline{D}_{n-1}(\varepsilon,\eta)+
ue^{-i\varepsilon}\overline{L}_{n-1}(\varepsilon,\eta)
+0 \overline{R}_{n-1}(\varepsilon,\eta)
$$
$$
\eqno(4.19)
$$
$$
\overline{R}_{n}(\varepsilon,\eta)=u\alpha
e^{i\varepsilon}\overline{U}_{n-1}(\varepsilon,\eta)+
u{ \bar{\alpha}}e^{i\varepsilon}\overline{D}_{n-1}(\varepsilon,\eta)
+0\overline{L}_{n-1}(\varepsilon,\eta)
+ue^{i\varepsilon}\overline{R}_{n-1}(\varepsilon,\eta)
$$
$$
\eqno(4.20)
$$ 
Call  ${\psi}_{n}(\epsilon,\eta)$ the matrix
$$
\psi_{n}=
\left(
\begin{array}{clcr}
\overline{U}_{n} & \overline{D}_{n} & \overline{L}_{n} & \overline{R}_{n} 
\end{array} \right)
\eqno(4.21)
$$
Then, from (4.17-20) we obtain that
$$
\psi_{n}(\varepsilon,\eta)=\psi_{n-1}(\varepsilon,\eta) uM, 
\eqno(4.22)
$$
where
$$
M=
\left(
\begin{array}{clcr}
v & 0 &\overline{\alpha h}  & \alpha h \\
0 & \bar{v} & \alpha \bar{h}  & \bar{\alpha} h \\
\alpha v & \overline{\alpha v} & \bar{h} & 0 \\
\bar{\alpha} v  & \alpha \bar{v} & 0 & h 
\end{array} \right)
\eqno(4.23)
$$
with $v=e^{-i\eta}$, $\overline{h}=e^{-i\varepsilon}$,
$h=e^{i\varepsilon}$, $\overline{v}=e^{i\eta}$ and $\alpha = 
e^{i \frac{\pi}{4}}$.

Call $ 1, 2, 3$ and $4$ the directions shown in the Figure 8 below:

\begin{picture}(250,165)(50,-40)
  \put(200,55){\line(1,0){50}}	
  \put(225,30){\line(0,1){50}}	
  \put(225,83){\vector(0,1){1}}
  \put(222,90){1}
  \put(222,15){2}
  \put(254,55){\vector(1,0){1}}
  \put(195,55){\vector(-1,0){1}}
  \put(184,53){4}
  \put(225,30){\vector(0,-1){1}}
  \put(265,53){3}
\put(100,-10){Figura 8. Directions associated to $M_{ij}$.}
\end{picture}

\noindent Notice that the subindices $i,j$ of $M_{ij}$ are in one-to-one
with the directions.
Indeed, $uM_{1j}$ corresponds to the amplitude of arrival at $(x,y)$ 
 ( in $(\epsilon,\eta)$
space)
coming up in the $(n-1)$-th step, $\forall j$, but going up if $j=1$,
down if $j=2$,
coming from the left if $j=3$ and coming from the right if $j=4$
in the $n$-th step.
Therefore, $uM_{1j}$ is the amplitude of arrival at 
$(x,y)$ ( in $(\epsilon,\eta)$ space ) following directions $1$ and $j$
in the $(n-1)$-th and $n$-th steps, respectively. More generally, $uM_{ij}$
is the amplitude of arrival at $(x,y)$ following directions $i$ and $j$ in the
$(n-1)$-th and $n$-th steps, respectively. 
\\
\\
\\
\noindent From now on only closed paths starting at (0,0) and
arriving at (0,0) in $n$ steps will be considered.

From
(4.22) it follows that
$$
\psi_{n}(\varepsilon,\eta)=\psi_{n-1}(\varepsilon,\eta)(uM)
=\psi_{n-2}(\varepsilon,\eta)(uM)^2
=\cdots =\psi_0 (uM)^{n}
\eqno(4.24)
$$
Denote by $ \psi_{0,i}$, $1\leq i\leq 4$, 
the line matrix with the only element
distinct from zero and equal to $1$ in the $i$-th column. 
Let $\psi_{0} \equiv \psi_{0,i}$ according to whether the path arrives
at the origin moving up ($i=1$), down ($i=2$), from the left ($i=3$) or 
from the right ($i=4$), respectively.
Then
$$               
\overline{F}_{n}(\varepsilon,\eta)=\psi_{0,i}(uM)^{n}\psi_{0,i}^T
\eqno(4.25)
$$
where $i=1,2,3,4$ if $F=U,D,L,R$, respectively, and $\Psi^{T}$ is the 
transpose of $\Psi$ and
$$
\sum_{F_{n} \in B_{n}} \overline{F}_{n}(\epsilon,\eta)=
\sum_{k=1}^{4}\psi_{0,k}
(\varepsilon,\eta)(uM)^{n}\psi_{0,k}^T(\varepsilon,\eta) 
\eqno(4.26)
$$
Given a $4 \times 4$ matrix $A$,
 $\psi_{0,i}A$  is the line matrix 
formed by the elements in the $i$-th line of $A$, that is,
$$
\psi_{0,i}A=
\left(
\begin{array}{clcr}
 A_{i,1} & A_{i,2} & A_{i,3} & A_{i,4}
\end{array} \right)
\eqno(4.27)
$$
so $\psi_{0,i}A\psi_{0,i}^T=A_{ii}$. Therefore, 
the sum over $i$ equals the trace of $A$. Thus,
$$
\sum_{k=1}^{4}\psi_{0,k}(uM)^{n}
\psi_{0,k}^T=Tr(uM)^{n}
\eqno(4.28)
$$
The total  partial amplitude of arrival at $(0,0)$ of  closed
paths moving in any direction in $n$ steps
given by (4.12) can be expressed compactly as
$$
\sum_{F_{n} \in B_{n}} F_{n}(0,0) 
\eqno(4.29)
$$
From (4.15), (4.26) and (4.28), it follows that
$$
 \sum_{F_{n} \in B_{n}} F_{n}(0,0)
=\int_0^{2\pi}\int_0^{2\pi}Tr(uM)^{n}\frac{d\varepsilon d\eta}{(2\pi)^2}
\eqno(4.30)
$$

To better understand relation (4.30),
consider the matrix $u^{n}M^n$, for some $n$. An element
 $(u^{n}M^n)_{i_1i_{n+1}}$ of this matrix is given as
$$
(u^{n}M^n)_{i_1i_{n+1}}=\sum_{i_2,\dots,i_n=1}^{4}uM_{i_1i_2} uM_{i_2i_3}\dots
uM_{i_{n-1}i_n} uM_{i_ni_{n+1}} .
\eqno(4.31)
$$ 
Recall that $uM_{i,j}$ is the partial amplitude of a path arriving at a site
coming from direction $i$ and going to the next site in one step following
direction $j$. Thus, each term in the r.h.s. of (4.31) is the amplitude
of a path of length $n$ starting at $P_{1}$ coming
from direction $i_{1}$, going to $P_{2}$
following direction $i_{2}$, etc, and
arriving  at site $P_{n+1}$ following 
direction $i_{n+1}$  after $n$ steps. The element $(u^{n}M^n)_{i_1i_{n+1}}$
gives  the total partial amplitude of arrival at $P_{n+1}$ in $n$ steps in
$(\epsilon,\eta)$ space. 

The terms in $(u^{n}M^{n})_{i_{1}i_{n+1}}$
describe open as well as closed paths. Let's see some examples.

\noindent{\bf Example 5.} Take
$n=5$, $i_1=2$ and $i_6=1$. 
The term $M_{23}M_{31}M_{11}M_{14}M_{41}$  describes a path
beginning at $P_{1}$ where it arrived coming from direction
$i_1=2$, going to $P_2$, $P_{3}$, $P_{4}$, $P_{5}$ and
to $P_{6}$ following directions $i_2=3$, $i_{3}=1$, $i_{4}=1$, $i_{5}=4$
and $i_{6}=1$, respectively. See Figure 9a) . 

\begin{picture}(300,170)(-170,10)
  \put(-100,50){\line(1,0){35}}
  \put(-100,50){\circle*{3}}
  \put(-120,50){$P_1$}
  \put(-82.54,50){\vector(1,0){1}}
  \put(-65,50){\circle*{3}}
  \put(-55,50){$P_2$}
  \put(-65,50){\line(0,1){70}}
  \put(-65,67.5){\vector(0,1){1}}
  \put(-65,85){\circle*{3}}
  \put(-65,102.5){\vector(0,1){1}}
  \put(-65,120){\circle*{3}}
  \put(-55,120){$P_4$}
  \put(-55,85){$P_3$}
  \put(-65,121){\line(-1,0){35}}
  \put(-82,121){\vector(-1,0){1}}
  \put(-100,121){\circle*{3}}
  \put(-120,121){$P_5$}
  \put(-100,121){\line(0,1){35}}
  \put(-100,138.5){\vector(0,1){1}}
  \put(-100,156){\circle*{3}}
  \put(-120,156){$P_6$}
  \put(-90,30){(a)}

  \put(90,50){\line(1,0){35}}
  \put(90,50){\line(0,1){70}}
  \put(90,120){\line(1,0){35}}
  \put(125,50){\line(0,1){70}}
  \put(90,50){\circle*{3}}
  \put(90,85){\circle*{3}}
  \put(125,85){\circle*{3}}
  \put(125,119){\circle*{3}}
  \put(90,119){\circle*{3}}
  \put(125,50){\circle*{3}}
  \put(70,50){$P_1$}
  \put(70,85){$P_6$}
  \put(70,120){$P_5$}
  \put(135,50){$P_2$}
  \put(135,85){$P_3$}
  \put(135,120){$P_4$}
  \put(125,102.5){\vector(0,1){1}}
  \put(90,100){\vector(0,-1){1}}
  \put(107.5,50){\vector(1,0){1}}
  \put(90,68){\vector(0,-1){1}}
  \put(125,68){\vector(0,1){1}}
  \put(107.5,120){\vector(-1,0){1}}
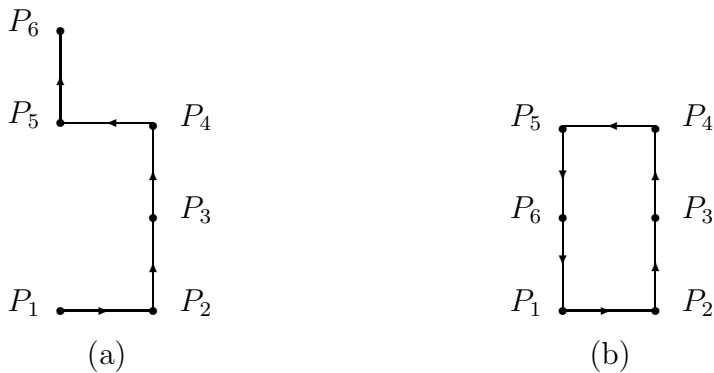
  \put(-110,15){Figure 9: Paths in (a) Ex. 5 and in (b) Ex. 6.}
  \put(100,30){(b)}
\end{picture}

\noindent{\bf Example 6.} Take $n=6$, $i_{1}=i_{6}=2$ and
the term $M_{23}M_{31}M_{11}M_{14}M_{42}M_{22}$
of $(M^{6})_{22}$ . This term describes the closed path in Fig. 9b.
\noindent The elements of $M^{n}$ outside the diagonal  have associated to 
them only  open paths. This is implied by the simple fact that these
elements have $i_{1} \neq i_{n+1}$. Closed paths are to be found only in the
diagonal elements since there $i_{1}=i_{n+1}$. However, open paths can also
be associated to some terms in the diagonal elements. Let's see some
examples.

\noindent{\bf Example 7}. Take $n=2$,  $i_{1}=i_{3}=1$ and the element 
$(M^2)_{11}=M_{11}M_{11}+M_{12}M_{21}+M_{13}M_{31}+M_{14}M_{41}$ with
$$
u^{2}(M^2)_{11}
=u^{2}v^2+0+u^{2}(\overline{\alpha}\overline{h})(\alpha v)+
u^{2}(\alpha h)(\overline{\alpha}v)
\eqno(4.32)
$$
To each one of the terms of $(M^{2})_{11}$ correspond the paths 
(a), (b), (c) and (d),
respectively, shown in  Figure 10.

\begin{picture}(170,170)(-120,-50)
  \put(-95,35){\line(0,1){50}}	
  \put(-95,35){\circle*{3}}
  \put(-80,80){$P_3$}	
  \put(-95,85){\circle*{3}}
  \put(-80,55){$P_2$}
  \put(-95,60){\circle*{3}}
  \put(-80,30){$P_1$}
  \put(-95,-15){(a)}
  \put(-95,73){\vector(0,1){1}}
  \put(-95,48){\vector(0,1){1}}
  \put(-95,15){\vector(0,1){10}}

  \put(15,85){\circle*{3}}
  \put(30,80){$P_1$}	
  \put(15,35){\circle*{3}} 
  \put(30,35){$P_2$}
  \put(15,35){\line(0,1){50}}	
  \put(15,-15){(b)} 
  \put(15,48){\vector(0,1){1}}
  \put(15,71){\vector(0,-1){1}}

  \put(210,50){\line(1,0){35}}	
  \put(245,50){\line(0,1){35}}	
  \put(228,50){\vector(1,0){1}}
  \put(245,68){\vector(0,1){1}}
  \put(210,50){\circle*{3}}
  \put(190,45){$P_1$}
  \put(255,45){$P_2$}
  \put(255,80){$P_3$}
  \put(245,50){\circle*{3}}
  \put(245,85){\circle*{3}}
  \put(210,30){\vector(0,1){10}}
  \put(235,-15){(d)}

 \put(110,50){\line(1,0){35}}	
  \put(110,50){\line(0,1){35}}
  \put(110,50){\circle*{3}}
  \put(85,80){$P_3$}
  \put(155,45){$P_1$}
  \put(85,45){$P_2$}
  \put(110,85){\circle*{3}}
  \put(145,50){\circle*{3}}
  \put(128,50){\vector(-1,0){1}}
  \put(110,68){\vector(0,1){1}}
  \put(145,30){\vector(0,1){10}}
  \put(125,-15){(c)}	

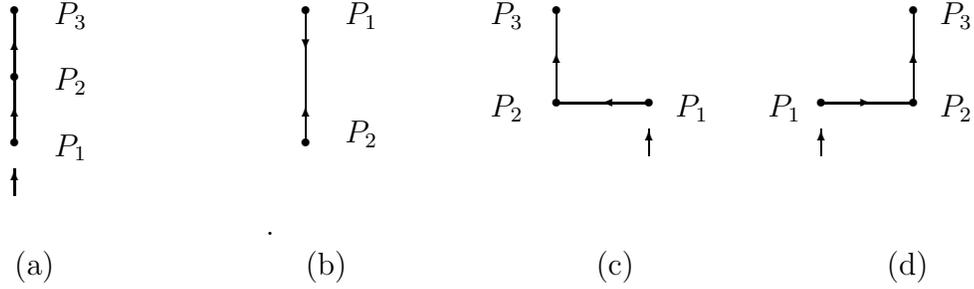
\put(-50,-40){Figure 10: Paths associated to the terms in $(M^{2})_{11}$}.
\end{picture}

\noindent{\bf Example 8.} Take $n=4$ 
and consider the following terms in  $(M^4)_{11}$:

\noindent a)The term $u^{4}M_{11}M_{11}M_{11}M_{11}=u^{4}v^4$ 
is the amplitude of  the open
path
shown in fig. 11 below. 

\begin{picture}(230,125)(-260,-10)

  \put(-95,35){\line(0,1){80}}	
  \put(-95,35){\circle*{3}}
  \put(-95,55){\circle*{3}}
  \put(-95,75){\circle*{3}}
  \put(-95,95){\circle*{3}}
  \put(-95,115){\circle*{3}}
  \put(-95,65){\vector(0,1){1}}
  \put(-95,45){\vector(0,1){1}}
  \put(-95,85){\vector(0,1){1}}
  \put(-95,105){\vector(0,1){1}}
\put(-150,-10){Figure 11: Path $(M_{11})^{4}$. }
\end{picture}

\noindent b) The term
$M_{11}M_{11}M_{13}M_{31}=vv(\overline{\alpha}\overline{h})(\alpha v)
=v^{3} \bar{h} v$
whose associated open path  is shown in Figure
 12.

\begin{picture}(280,145)(-180,-40)
  \put(-8,-10){\line(0,1){70}}
  \put(-8,60){\line(1,0){35}}
  \put(27,60){\line(0,1){35}}
  \put(-8,-10){\circle*{3}}
  \put(-8,25){\circle*{3}}
  \put(-8,60){\circle*{3}}
  \put(27,60){\circle*{3}}
  \put(27,95){\circle*{3}}
  \put(-8,7.5){\vector(0,1){1}}
  \put(-8,42.5){\vector(0,1){1}}
  \put(9.5,60){\vector(1,0){1}}
  \put(27,77.5){\vector(0,1){1}}
  \put(-28,-10){$P_1$}
  \put(-28,25){$P_2$}
  \put(-28,60){$P_3$}
  \put(37,60){$P_4$}
  \put(37,95){$P_5$}
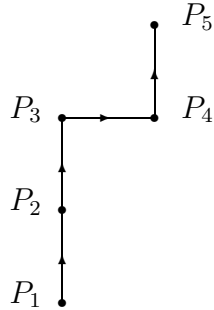
  \put(-90,-30){Figure 12: Path $M_{11}M_{11}M_{13}M_{31}$.}

\end{picture}

\noindent c) The term $u^{4}M_{13}M_{32}M_{24}M_{41}=u^{4}
(\overline{\alpha}\overline{h})
(\overline{\alpha}\overline{v})(\overline{\alpha}h)(\overline{\alpha}v)
=u^{4}\overline{\alpha}^4
(\overline{h}h)(\overline{v}v)=u^{4}\overline{\alpha}^4$
is the amplitude of the closed path shown in Fig. 13:

\begin{picture}(100,100)(-200,5)
  \put(-40,50){\line(1,0){35}}	
  \put(-40,50){\line(0,1){35}}
  \put(-40,85){\line(1,0){35}}
  \put(-5,50){\line(0,1){35}}
  \put(-5,85){\circle*{3}}
  \put(-40,50){\circle*{3}}
  \put(5,80){$P_2$} 
  \put(5,45){$P_1$}
  \put(-65,45){$P_4$}
  \put(-65,80){$P_3$}
  \put(-40,85){\circle*{3}}
  \put(-5,50){\circle*{3}}
  \put(-22,50){\vector(1,0){1}}
  \put(-40,68){\vector(0,-1){1}}
  \put(-5,68){\vector(0,1){1}}
  \put(-22,85){\vector(-1,0){1}}
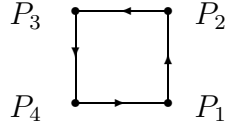
  \put(-100,5){Figure 13: Path $M_{13}M_{32}M_{24}M_{41}$. }
\end{picture}

In order to restrict to the elements of $M^{n}$ having closed paths we
must take the trace of $M^{n}$.
A closed path begins at and return to 
$P_{1}$
after $n$ steps. Since it is closed it has to cover $n/2$ horizontal
bonds in one direction and $n/2$ horizontal bonds following the opposite
direction. 
The same is true for the vertical bonds traversed by $p$.
So, if the term
$
M_{i_1i_2}M_{i_2i_3}\dots
M_{i_{n-1}i_n}M_{i_ni_{n+1}} 
$ 
, $i_{n+1}=i_{n}$, describes a closed path, then
the number of $h$'s ($v$'s) equals the number of $\bar h$'s ($\bar v$'s)
 appearing in it. 
In this case,  it's possible to organize the term into
a product of pairs $h \bar h =1$ and $v \bar v =1$ and the double
integral in $\epsilon$ and $\eta$ will give $(2\pi)^{2}$ times a
product of $\alpha$'s and $\bar \alpha$'s. More precisely, the
double integral over a closed path in $Tr M^{n}$ equals
$$
(2\pi)^2 \prod \alpha \prod \overline{\alpha}
\eqno(4.33)
$$
where the first product is over all counterclockwise rotations
and the second is over all clockwise rotations, so
$$
\prod \alpha \prod \overline{\alpha}  =(-1)^{t(p)}
\eqno(4.34)
$$
where $t(p)$ is the number of complete  $2\pi$ revolutions performed
by a tangent vector traversing the closed path $p$.
Remind that one has yet to multiply (4.34) by $(-1)$ in order to get 
the complete sign $s(p)$ of $p$.

If a path is open the $h$'s, $\bar{h}$'s ($v$'s, $\bar{v}$'s) don't
match up into pairs.
There will be left integrals of the form
$$
\int_{0}^{2\pi}e^{i k \theta}d\theta =0,
\eqno(4.35)
$$
where $\theta$ stands for $\eta$ or $\epsilon$
and $k \geq 1$, hence,
the  integrals in $\eta$ and $\varepsilon$ remove completely 
terms describing open
paths. Let's see examples.

\noindent{\bf Example 9.} Take $n=1$. In this case there are only open paths
and
$$
\int_0^{2\pi} \int_0^{2\pi}d\eta d\varepsilon M_{ij}=0,\qquad \quad \forall i,j
\eqno(4.36)
$$
\noindent{\bf Example 10.}
Using (4.35), in ex. 7,
$$
\int_{0}^{2\pi} \int_{0}^{2\pi}  d\eta d\varepsilon (M^2)_{11}=0
\eqno(4.37)
$$
\noindent{\bf Example 11.} It is clear that
$$
\int d\varepsilon d\eta (M^n)_{ij}=0\qquad \quad \forall i,j
\eqno(4.38)
$$
if $n=1,2,3$, 
which is guaranteed by the fact that
in a square lattice  closed paths
are possible only if $n\geq 4$.
In the case $n=2$ the
path in Figure 10.b) 
is closed but it traverses the same edge back and its amplitude is
thus zero.

\noindent{\bf Example 12.} Using (4.35) in ex.8, 
for the term $M_{11}M_{11}M_{11}M_{11}=v^{4}$
$$
\int_0^{2\pi} \int_0^{2\pi}v^4d\eta d\varepsilon =2\pi\int_0^{2\pi}e^{-4i\eta}d\eta=0.
\eqno(4.39)
$$
For the term $M_{11}M_{11}M_{13}M_{31}=v^{3} \bar h$, 
$$
\int_0^{2\pi}d\eta d\varepsilon v^{3} \bar h =0
\eqno(4.40)
$$
For the term 
$M_{13}M_{32}M_{24}M_{41}=\overline{\alpha}^4$
which describes a closed path the result
$$
\frac{1}{(2\pi)^2}\int_0^{2\pi}d\eta d\varepsilon
\overline{\alpha}^4=\overline{\alpha}^4.
\eqno(4.41)
$$
follows which has the form (4.33-34) with $\bar{\alpha}^{4} \equiv 1$.

Given a closed path in
 $(M^n)_{i i}$, the inverse path is present
in some $(M^n)_{jj}$, $j\ne i$. 
For instance, in
 $(M^4)_{11}$ there are the closed paths shown in Fig. 14
given by the terms $M_{14}M_{42}M_{23}M_{31}$ and $M_{13}M_{32}M_{24}M_{41}$.

\begin{picture}(280,145)(-120,-40)
  \put(-40,50){\line(1,0){35}}	
  \put(-40,50){\line(0,1){35}}
  \put(-40,85){\line(1,0){35}}
  \put(-5,50){\line(0,1){35}}
  \put(-5,85){\circle*{3}}
  \put(5,80){$P_1$} 
  \put(-22,50){\vector(1,0){1}}
  \put(-40,68){\vector(0,-1){1}}
  \put(-5,68){\vector(0,1){1}}
  \put(-22,85){\vector(-1,0){1}}
  \put(-28,30){(a)}

 \put(110,50){\line(1,0){35}}	
  \put(110,50){\line(0,1){35}}
  \put(110,85){\line(1,0){35}}
  \put(145,50){\line(0,1){35}}
  \put(110,85){\circle*{3}}
  \put(85,80){$P_1$} 
  \put(128,50){\vector(-1,0){1}}
  \put(110,68){\vector(0,1){1}}
  \put(145,68){\vector(0,-1){1}}
  \put(128,85){\vector(1,0){1}}
  \put(122,30){(b)}
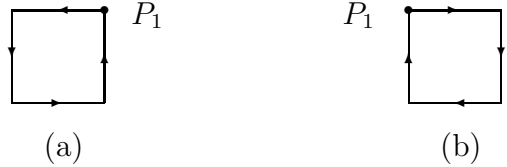
  \put(-110,-10){Figure 14: Paths (a) $M_{14}M_{42}M_{23}M_{31}$ and 
(b) $M_{13}M_{32}M_{24}M_{41}$.}
\end{picture}

\indent In $(M^4)_{22}$ there are the terms
$M_{24}M_{41}M_{13}M_{32}$ and  $M_{23}M_{31}M_{14}M_{42}$
, with associated closed paths  shown in Fig. 15c and 15d,
respectively:

\begin{picture}(280,145)(-120,-40)
  \put(-40,50){\line(1,0){35}}	
  \put(-40,50){\line(0,1){35}}
  \put(-40,85){\line(1,0){35}}
  \put(-5,50){\line(0,1){35}}
  \put(-5,50){\circle*{3}}
  \put(5,45){$P_1$} 
  \put(-22,50){\vector(-1,0){1}}
  \put(-40,68){\vector(0,1){1}}
  \put(-5,68){\vector(0,-1){1}}
  \put(-22,85){\vector(1,0){1}}
  \put(-28,30){(c)}

 \put(110,50){\line(1,0){35}}	
  \put(110,50){\line(0,1){35}}
  \put(110,85){\line(1,0){35}}
  \put(145,50){\line(0,1){35}}
  \put(110,50){\circle*{3}}
  \put(85,45){$P_1$} 
  \put(128,50){\vector(1,0){1}}
  \put(110,68){\vector(0,-1){1}}
  \put(145,68){\vector(0,1){1}}
  \put(128,85){\vector(-1,0){1}}
  \put(122,30){(d)}
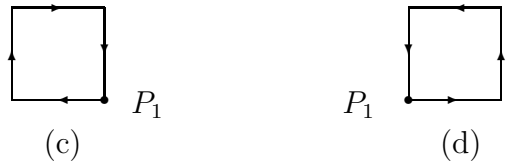
  \put(-110,-10){Figura 15: Paths (c) $M_{24}M_{41}M_{13}M_{32}$ and 
(d) $M_{23}M_{31}M_{14}M_{42}$}
\end{picture}

\indent In $(M^4)_{33}$, there are the terms 
$M_{32}M_{24}M_{41}M_{13}$ and $M_{31}M_{14}M_{42}M_{23}$
with associated closed paths shown in Figure 16e) and 16f), respectively.

\begin{picture}(280,145)(-120,-40)
  \put(-40,50){\line(1,0){35}}	
  \put(-40,50){\line(0,1){35}}
  \put(-40,85){\line(1,0){35}}
  \put(-5,50){\line(0,1){35}}
  \put(-5,85){\circle*{3}}
  \put(-1,80){$P_1$} 
  \put(-22,50){\vector(-1,0){1}}
  \put(-40,68){\vector(0,1){1}}
  \put(-5,68){\vector(0,-1){1}}
  \put(-22,85){\vector(1,0){1}}
  \put(-28,30){(e)}

  \put(110,50){\line(1,0){35}}	
  \put(110,50){\line(0,1){35}}
  \put(110,85){\line(1,0){35}}
  \put(145,50){\line(0,1){35}}
  \put(145,50){\circle*{3}}
  \put(150,45){$P_1$} 
  \put(128,50){\vector(1,0){1}}
  \put(110,68){\vector(0,-1){1}}
  \put(145,68){\vector(0,1){1}}
  \put(128,85){\vector(-1,0){1}}
  \put(122,30){(f)}
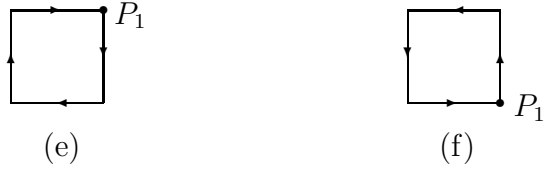
  \put(-110,-10){Figura 16. Paths (e) $M_{32}M_{24}M_{41}M_{13}$ and 
(f) $M_{31}M_{14}M_{42}M_{23}$ }
\end{picture}

\indent In $(M^4)_{44}$, there are the terms
 $M_{42}M_{23}M_{31}M_{14}$ and $M_{41}M_{13}M_{32}M_{24}$
with the associated
closed paths shown in Figure 17g) and 17h), respectively.

\begin{picture}(280,145)(-120,-40)
  \put(-40,50){\line(1,0){35}}	
  \put(-40,50){\line(0,1){35}}
  \put(-40,85){\line(1,0){35}}
  \put(-5,50){\line(0,1){35}}
  \put(-40,85){\circle*{3}}
  \put(-60,80){$P_1$} 
  \put(-22,50){\vector(1,0){1}}
  \put(-40,68){\vector(0,-1){1}}
  \put(-5,68){\vector(0,1){1}}
  \put(-22,85){\vector(-1,0){1}}
  \put(-28,30){(g)}

 \put(110,50){\line(1,0){35}}	
  \put(110,50){\line(0,1){35}}
  \put(110,85){\line(1,0){35}}
  \put(145,50){\line(0,1){35}}
  \put(110,50){\circle*{3}}
  \put(90,50){$P_1$} 
  \put(128,50){\vector(-1,0){1}}
  \put(110,68){\vector(0,1){1}}
  \put(145,68){\vector(0,-1){1}}
  \put(128,85){\vector(1,0){1}}
  \put(122,30){(h)}
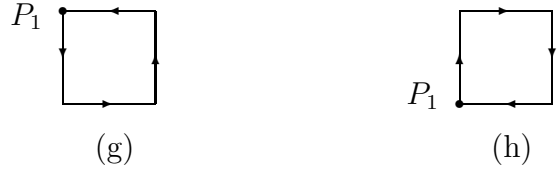
  \put(-110,-10){Figura 17. Paths (g) $M_{42}M_{23}M_{31}M_{14}$ and 
(h) $M_{41}M_{13}M_{32}M_{24}$}
\end{picture}

Note that (e) is the inversion of (a), (f) is the inversion of (c), 
(g) of (b), and (h) of (d).

So restricting 
to the diagonal terms of $M^{n}$ which amounts to take the trace of this
matrix and then performing a double integration on the angles
to eliminate open paths, dividing the result by 2 to eliminate inversions,
and multiplying the result by $-u^{n}$ gives the total complete
amplitude ( with the right signs )
to arrive back at $P_{1}$ in $n$ steps moving in any direction. We have thus achieved the following
relation:
$$
\sum_{p(n,P_1)} W_{p}(u) =
-\frac{1}{2}\frac{1}{(2\pi)^2}\int_0^{2\pi}\int_0^{2\pi}d\varepsilon d\eta
Tr (uM)^n
\eqno(4.42)
$$
The above result is restricted to a fixed site $P_{1}$.
For the finite $N \times N$ lattice
 with $N^2$ sites and disregarding boundary effects, the total 
(independent of site) amplitude of closed paths of length $n$ is:
$$
N^2\sum_{p(n,P_1)}W_{p}(u) \equiv \sum_{p(n)}W_{p}(u)=
-\frac{N^2}{2}\frac{1}{(2\pi)^2}\int_0^{2\pi}\int_0^{2\pi}d\varepsilon d\eta
Tr(uM)^n
\eqno(4.43)
$$
Taking all $N^{2}$ sites into account imply that given a closed path
$p(n)$, the summation $\sum_{p(n)}W(p)$ includes
all circular permutations of $p$. To eliminate these the previous relation
has to be divided by
 $n$. Then, the amplitude is given by
$$
\frac{1}{n}\sum_{p(n)}W_{p}(u) =
-\frac{N^2}{2}\frac{1}{(2\pi)^2}\int_0^{2\pi}\int_0^{2\pi}d\varepsilon d\eta
\frac{Tr(uM)^n}{n}
\eqno(4.44)
$$
We notice that a nonperiodic path appears  $n$ times in the sum but a
periodic path of length $n$ and period $w$ has $n/w$ distinct starting
points only and for this reason it appears $n/w$ times in the sum over paths.
For instance, the periodic path $(D_{j_{1}}D_{j_{2}})(D_{j_{1}}D_{j_{2}})
(D_{j_{1}}D_{j_{2}})$ of length $n=6$ and period $w=3$ has only two
distinct starting points. The other equivalent periodic path
is $D_{j_{2}}(D_{j_{1}}D_{j_{2}})(D_{j_{1}}D_{j_{2}})D_{j_{1}}$.
After division by $n$, periodic paths with period $w$ will show up in
the sum with a weight $1/w$. Thus,
the above relation includes all closed paths of length
$n$
over the $N\times N$ lattice, periodic and nonperiodic, and excludes
inversions and circular permutations. 
The total amplitude 
of closed paths of any length is then given by the series
$$
\sum_{n} \frac{1}{n}\sum_{p(n)}W_{p}(u)
=\sum_{n=1}^{\infty}-\frac{N^2}{2}
\frac{1}{(2\pi)^2}\int_0^{2\pi}\int_0^{2\pi}d\varepsilon d\eta
\frac{Tr(uM)^n}{n} 
\eqno(4.45)
$$
whose convergence will be investigated below. 
We note that since the lattice is square, closed paths with nonzero amplitude
are possible only for $n > 3$ but in view of relation (4.38) in Ex. 11
we
can write the series in (4.45) starting from $n=1$.

With the above remarks,
$$
\sum_{n=1}^{\infty}\frac{1}{n}\sum_{p(n)} W_{p}(u)
\equiv \sum_{[p]}[ W_{p}(u)-\frac{1}{2}(W_{p}(u))^{2}
+\frac{1}{3}(W_{p}(u))^{3}-...]
\eqno(4.46)
$$
In $\sum_{[p]}$ the first term is the sum of $W_{p}(u)$ over all nonperiodic paths.
The other terms give the sum over all periodic paths
since any periodic path is the repetition of some nonperiodic path $p$
with period given by $w=2,3,...$.  
In section 3.2 the sign of a periodic
path was proved to be $-1$ if $w$ is even and equal to the sign 
of its nonperiodic subpath
if $w$ is odd. This explains the signs in the r.h.s of (4.46).

Since $|u| < 1$ then $|W|<1$ and the series between brackets converges to
$ln(1+W)$, and
the r.h.s. of (4.46) equals to
$$
\sum_p \ln(1+W_{p}(u))=\ln \prod_p (1+W_{p}(u)).
\eqno(4.47)
$$ 
a result to be used below.

\noindent{\bf Theorem 4.1.} Take $|u| \leq r < \frac{1}{4}$. Then, the series 
$$
\sum_{n=1}^{\infty}\frac{u^{n}(M^{n})_{ij}}{n}
\eqno(4.48)
$$
converges uniformily.

\noindent{\bf Proof:}
We have that $ |M_{ij}| \leq 1 $, $\forall i,
j=1,2,3,4$, so from (4.31) we get
$$
\mid (M^{n})_{ij} \mid  \leq 4^{n-1}
\eqno(4.49)
$$
and
$$
\left| \frac{u^{n}(M^{n})_{ij}}{n} \right|  \leq \frac{ 4^{n-1}
\mid u \mid^{n}}{n}
\eqno(4.50)
$$
The series
$$
\sum_{n=1}^{\infty} \frac{(4 \mid u \mid)^{n}}{n}
\eqno(4.51)
$$
converges for $|u| \leq r < \frac{1}{4}$, hence, 
by Weierstrass $M$-test the series
(4.48) converges uniformly for $|u| \leq r <1/4$. 

We may conclude that the series
$\sum_{n=1}^{\infty} \frac{ -(u M)^{n}}{n}$
converges uniformly  to the matrix $ln(1-uM)$
in the same interval.

We may now integrate the series term by term to get the series (4.45)
which likewise converges uniformly in the same interval.
Interchanging integration and summation in (4.45)
yields
$$
\frac{N^2}{2}
\frac{1}{(2\pi)^2}\int_0^{2\pi}\int_0^{2\pi}d\varepsilon d\eta
Tr \sum_{n=1}^{\infty}-\frac{(uM)^n}{n}= 
\frac{N^2}{2}
\frac{1}{(2\pi)^2}\int_0^{2\pi}\int_0^{2\pi}d\varepsilon d\eta
Tr ln( I -uM) 
\eqno(4.52)
$$
From the previous analysis, $|u| \leq r <1/4$. However,
 the r.h.s. of (4.52) is well defined in a bigger
domain. Using the relation
$$
Tr ln(I-uM)=ln det(I-uM)
\eqno(4.53)
$$
which is valid for $det(1-uM) \neq 0$ [18],
we get
$$
\frac{N^2}{2}
\frac{1}{(2\pi)^2}  \int_0^{2\pi}\int_0^{2\pi}d\varepsilon d\eta
lndet( I -uM)
\eqno(4.54)
$$ 
The determinant can be easily computed and one finds that
$$
\det(1-uM)=(u^2+1)^2-2u(1-u^2)(\cos \varepsilon+\cos \eta)
\eqno(4.55)
$$

\noindent Taking the logarithm on both sides of relation (3.24)
gives
$$
\frac{ln Z(u)}{N^{2}}=ln2+2(1-\frac{1}{N})ln(coshK)
 + ln\prod_{p}[1+W_{p}(u)]
\eqno(4.56)
$$
or, using (4.45-47), (4.52,4.54),
$$
\frac{ln Z(u)}{N^{2}} = ln2 + 2(1-
\frac{1}{N})ln(coshK)
+\frac{1}{8 {\pi}^{2}}\int_{0}^{2\pi}\int_{0}^{2\pi} d\varepsilon d\eta
ln det(1-uM)
\eqno(4.57)
$$
Using (4.55) with $u=tanh K$ and the relations
$$
u=\frac{1}{2}sinh(2K)(1-u^{2})
\eqno(4.58)
$$
$$
(1+u^{2})^{2}=cosh^{2}(2K) (1-u^{2})^{2}
\eqno(4.59)
$$
$$
(1-u^{2})=\frac{1}{cosh^{2}2K}
\eqno(4.60)
$$
gives
$$
det(1-uM)=cosh^{-4}2K \left[ (cosh2K)^{2} - (sinh 2K)
(cos \eta + cos \varepsilon) \right]
\eqno(4.61)
$$

\noindent {\bf Onsager's formula} follows from (4.57) after taking
the limit $N \rightarrow \infty$:
$$
-\frac{f}{k_{B}T} =
ln2  +\frac{1}{2 {\pi}^{2}}\int_{0}^{\pi}\int_{0}^{\pi} d\varepsilon d\eta
ln \left[ (cosh2K)^{2} - sinh 2K
(cos \eta + cos \varepsilon) \right]
\eqno(4.62)
$$
where $f$ is the free energy per
site in the thermodynamical limit ( see (2.5)).

The integral in (4.62) can not be evaluated in terms
of simple functions. The derivatives of the integral however
can be expressed in terms of elliptic functions [2,15,19].

Set $2k=tanh 2K/ cosh 2K$. Then,
$$
-\frac{f}{k_{B}T} =
ln2  + ln(cosh2K) + \frac{1}{2 {\pi}^{2}}\int_{0}^{\pi}\int_{0}^{\pi} d\varepsilon d\eta
ln \left[ 1 -  2k
(cos \eta + cos \varepsilon) \right]
\eqno(4.63)
$$
Expanding the logarithm in powers of $k$  it follows that
$$
-\frac{f}{k_{B}T}= ln2 + ln(cosh 2K) - \sum_{n=1}^{\infty}
\left( \frac{(2n)!}{(n!)^{2}} \right)^{2} k^{2n}
\eqno(4.64)
$$
The series converges for $|2k(\cos \epsilon +\cos \eta)| \leq 4 |k| <1$.
For $k > 0$ ($J>0$) and at $k=1/4$, that is,
 at the critical value $K=K_{c}$ ( or
temperature $T_{c}= 2J k_{B}^{-1} ln^{-1}(\sqrt{2} +1)$ )
given by
$$
2senh 2K_{c} = cosh^{2} 2K_{c}
\eqno(4.65)
$$ it diverges. ( Similarly, for $J < 0$ which implies $k < 0$ 
and divergence at $T_{c}= 2J k_{B}^{-1} ln^{-1}(\sqrt{2} - 1)$ ).

The internal energy $U$ is given by
$$
U = -kT^{2} \frac{\partial}{\partial T} \frac{f}{kT}
\eqno(4.66)
$$
From (4.63),
$$
U= -J coth(2K) \left[ 1+ (sinh^{2}2K -1) I(2K) \right]
\eqno(4.67)
$$
where
$$
I(2K)= \frac{1}{\pi ^{2}} \int_{0}^{\pi} \frac{d \epsilon d \eta}{
cosh^{2}(2K) - sinh(2K)(cos \epsilon + cos \eta)}
\eqno(4.68)
$$
By performing one of the integrals its found that
$$
U= -J coth(2K) \left[ 1+ (2 tanh^{2}2K -1) \frac{2}{\pi} F(k_{1}) \right]
\eqno(4.69)
$$
where $k_{1}=4k$
and $F(k_{1})$ is the complete elliptic integral of the first kind 
defined by
$$
F(k_{1})= \int_{0}^{\pi /2} (1 - k_{1}^{2} sen^{2}\theta)^{-\frac{1}{2}}
d \theta
\eqno(4.70)
$$
The elliptic function $F$ ( see Ref. [19] ) has the property that 
$$
F \rightarrow ln [4(1-k_{1}^{2})]^{-1/2}
\eqno(4.71)
$$
as $k_{1} \rightarrow 1^{-}$. So it diverges logarithmically at $k_{1}=1$,
or at the value $K_{c}$ given by (4.65).

In relation (4.69) for $U$ the function $F$ is multiplied by $(2tanh^{2}2K-1)$ which is zero
at the critical point $K_{c}$. 
Indeed, 
from the identity $cosh^{2}x = 1+sinh^{2}x$ relation (4.65) implies that
$sinh(2K_{c})=1$. Using this and (4.65), $tanh^{2}2K_{c} =1/2$ follows.
So the function $U$ is  continuous
at $K_{c}$.

The specific heat can be computed from the definition
$$
C=\frac{\partial U}{\partial T}
\eqno(4.72)
$$
It is given by
$$
C=\frac{2k_{1}}{\pi}(Kcoth 2K)^{2} \{ 2F(k_{1})-2E(k_{1}) 
+2(tanh^{2}2K-1)G(k_{1}) \}
\eqno(4.73)
$$
where
$$
G(k_{1})= \frac{\pi}{2} +(2tanh^{2}2K-1)F(k_{1})
\eqno(4.74)
$$
and $E(k_{1})$ is the complete elliptic integral of the second kind, defined
by
$$
E(k_{1})= \int_{0}^{\pi /2} (1 - k_{1}^{2} sen^{2}\theta)^{\frac{1}{2}}
d \theta
\eqno(4.75)
$$
which is well defined at $k_{1}=1$.
From the exact result (4.73) it follows that the specific heat is
logarithmically divergent at the critical point.


\newpage
\begin{thebibliography}{99}

\bibitem{1} E. Ising, Zeitschrift f. Physik {\bf 31}, 253 (1925).

\bibitem{2} L. Onsager, Phys. Rev. {\bf65}, 117 (1944).

\bibitem{3} M. Kac and J. C. Ward, Phys. Rev. {\bf 88}, 1332 (1952).

\bibitem{4} S. Sherman, J. Math. Phys. {\bf 1}, 202 (1960).

\bibitem{5} S. Sherman, Bull. Am. Math. Soc. {\bf 68}, 225 (1962).

\bibitem{6} S. Sherman, J. Math. Phys. {\bf 4}, 1213 (1963).

\bibitem{7} P. N. Burgoyne, J. Math. Phys. {\bf 4}, 1320 (1963).

\bibitem{8} G. A. T. F. da Costa, J. Math. Phys. {\bf 38}, 1014
(1997).

\bibitem{9} R. P. Feynman,  Statistical Mechanics. A set of lectures.
The Benjamin and Cummings Publishing Co., 1972. 

\bibitem{10} H. S. Green and C. A.  Hurst, Order-Disorder Phenomena,
John Wyley and Sons.

\bibitem{11} S. G. Brush, Rev. Mod. Phys. ,{\bf39}, 883 (1967). 


\bibitem{12} N. V. Vdovichenko, Soviet Phys. JETP, {\bf 20}, 477 (1965).


\bibitem{13} Landau and Lifschitz, Statistical Physics, Addison-Wesley,
1969.


\bibitem{14} M. L. Glasser, Am. J. Phys., {\bf 38}, 1033, 1970. 

\bibitem{15} C. J. Thompson, Mathematical Statistical Mechanics, Princenton
University Press, 1972.

\bibitem{16} G. F. Newell and E. W. Montroll, Rev. Mod. Phys. , {\bf25}, 
353 (1953).


\bibitem{17} B. A. Cipra, The American Mathematical Monthly, {\bf  94}, 
,
937 (1987).

\bibitem{18} F. Brauer and J. Nohel, The qualitative Theory of ODE,
W. A. Benjamin, INC., 1969.

\bibitem{19} M. Abromowitz and I. A. Stegun, Handbook of Mathematical
Functions, Dover, 1972.


\end{thebibliography}
\end{document}